\documentclass[prd,showpacs,amsmath,amssymb,superscriptaddress,nofootinbib,showkeys]{revtex4}
\usepackage{amssymb}
\usepackage{amsmath}
\usepackage[cp1251]{inputenc}
\usepackage[T2A]{fontenc}
\usepackage{graphicx}
\usepackage{floatflt}
\DeclareGraphicsExtensions{.pdf,.eps,.png,.jpg}

\begin{document}
\title{The nonlinear anisotropic model of the Universe with the linear potential}

\author{Ruslan K. Muharlyamov}
\email{rmukhar@mail.ru} \affiliation{Department of General
Relativity and Gravitation, Institute of Physics, Kazan Federal
University, Kremlevskaya str. 18, Kazan 420008, Russia}

\author{Tatiana N. Pankratyeva}
\email{ghjkl.15@list.ru} \affiliation{Department of Higher
Mathematics, Kazan State Power Engineering University,
Krasnoselskaya str. 51, Kazan 420066, Russia}


\begin{abstract}
In the Bianchi I cosmology  some subclasses of the Horndeski theory allow for the non-standard
anisotropy behavior. For
example, the anisotropy are damped near the initial singularity
instead of tending to infinity. In this article, we analyze the
nonlinear anisotropic model with the linear potential.  We have
considered an example of such theory, for which the anisotropy is always finite.
The anisotropy  reaches its a maximum value at the initial moment. The anisotropy suppression occurs during the inflationary stage, and it approaches zero at later times. This cosmological model does not contain the
singular point.
\end{abstract}

\pacs{04.20.Jb}

\keywords{Horndeski theory, Bianchi I cosmology, dark energy}

\maketitle

\section{Introduction}

 It is usually stated that the state of the  modern Universe  is isotropic \cite{Rahman}. In his work \cite{Misner} Mizner C. W
speaks about this mysterious fact.
 Hawking and others formulated  the so-calledcosmic no-hair conjecture  long ago \cite{Gibbons, Hawking0}.
The states that the late  Universe is homogeneous and isotropic, i.e., would obey the cosmological principle, regardless of initial states of the Universe. Moreover the initial states of the Universe may not obey this principle. There are only partial proofs for this hypothesis, for example \cite{East, Carroll}. On the other hand in the work \cite{Colin}, based on SNe Ia datasets, the arguments are given for the anisotropic cosmic acceleration. Authors \cite{Zhao} examined the Pantheon sample of SNe Ia and five combinations among Pantheon. Authors conclude that the cosmic anisotropy is preferable for most observations. However  the cosmic anisotropy found in Pantheon sample significantly relies on the inhomogeneous distribution of SNe Ia in the sky. More homogeneous distribution of SNe Ia is necessary  for a clear answer. Thus the absolute isotropy of the modern Universe raises questions.

Was there an anisotropic phase in the past? The isotropy of the
Universe in the past does not follow from any general principles.
One of the main arguments for the existence of the anisotropic
phase of the Universe is the anisotropy at various scales of the
microscopic wave background (CMB). The CMB contains information
about the past of the Universe. There are anomalies in the
 CMB at the largest scales \cite{Smoot, Bennett, Hinshaw1,
Hinshaw2, Nolta, Hinshaw3, Ade, PlanckColl}. The Bianchi Universe can explain these
anomalies of the CMB \cite{Ellis, Komatsu, Thorsrud}.  The Bianchi type-I (BI) space-time is a popular model. BI  models have been studied from different perspectives \cite{Amirhashchi0,Amirhashchi,Nguyen,Koussour,Koussour1,Akarsu,Akarsu1,Sarmah,Hawking,Campanelli,Hu}.
For example, authors \cite{Akarsu,Akarsu1,Sarmah} consider constraints on the BI spacetime extension of the standard $\Lambda$CDM model.

An important criterion for the viability of any anisotropic model
is the dynamic properties  of anisotropic characteristics. In particular, it was argued in \cite{Hawking} that, from the point of view of the particle production, the isotropization of the Universe should occur quite early, no later than the beginning of primary nucleosynthesis  ($t\sim 1\,s$).  The work  \cite{Campanelli} analyzed the effects caused by cosmic anisotropy on the primordial production of $^4$He. It was found that in the anisotropic Universe there is an overproduction of $^4$He with respect to the standard isotropic case. In order to agree with  observational data, it is necessary to limit the anisotropy level at the time of freeze-out.
There is an opposite effect \cite{Hu}. Authors showed that the particle
production provides the early isotropization.  In General Relativity (GR) the process of isotropization occurs naturally.
When the Universe expands the anisotropy terms decrease faster
than the contribution of other forms of energy subject in the Einstein equations and the
Universe rapidly approaches a locally isotropic state
\cite{Jacobs}.
 The anisotropic terms become dominant when approaching the beginning of the Universe, then they  endure an infinite discontinuity.
The state of affairs changes for the modified theories
of gravity.  The issue of isotropization in the modified theories of gravity is important and is considered by many researchers \cite{Bhattacharya,Mandal,Alfedeel,Tahara,Sushkov,Arora,Sharif,Esposito}.
In the  BI cosmology, some subclasses of the Horndeski gravity (HG) allow for the
non-standard behavior of anisotropy . In the works \cite{Sushkov, SushkovStar, Muharlyamov0},
the HG theories were considered, in which the
anisotropy has nonmonotonic dynamics.

The HG is determined by the following action density
\cite{Horndeski}:
\begin{eqnarray}\label{action} L_H=\sqrt{-g}\Big(\mathcal{L}_2+\mathcal{L}_3+\mathcal{L}_4+\mathcal{L}_5\Big) \,,\end{eqnarray}
with functions (in the parameterization of ref.\cite{Kobayashi1}):
$$\mathcal{L}_2 = G_2(\phi,X)\,,\, \mathcal{L}_3 = -
G_3(\phi,X)\Box\phi\,,$$
$$\mathcal{L}_4 = G_{4}(\phi,X) R +G_{4X}(\phi,X) \left[ (\square
\phi )^{2}-(\nabla_\mu \nabla_\nu \phi)^2  \right]\,,$$
\begin{equation} \mathcal{L}_5 = G_{5}(\phi,X) G_{\mu\nu}\,\nabla^\mu \nabla^\nu
\phi -\frac{1}{6}G_{5X}   \left[\left( \Box \phi \right)^3 -3 \Box
\phi (\nabla_\mu \nabla_\nu \phi)^2 + 2\left(\nabla_\mu \nabla_\nu
\phi \right)^3 \right], \label{lagr2}
\end{equation}
respectively, where $g$ is the determinant of a metric tensor
$g_{\mu\nu}$; $R$ is the Ricci scalar and $G_{\mu\nu}$ is the
Einstein tensor; the factors $G_{i}$ ($i=2,3,4,5$) are arbitrary
functions of the scalar field $\phi$ and the canonical kinetic
term, $X=-\frac{1}{2}\nabla^\mu\phi \nabla_\mu\phi$. We consider
the definitions $G_{iX}\equiv \partial G_i/\partial X$,
 $(\nabla_\mu \nabla_\nu \phi)^2\equiv\nabla_\mu \nabla_\nu \phi
\,\nabla^\nu \nabla^\mu \phi$, and $\left(\nabla_\mu \nabla_\nu
\phi \right)^3\equiv \nabla_\mu \nabla_\nu \phi  \,\nabla^\nu
\nabla^\rho \phi \, \nabla_\rho \phi  \nabla^\mu \phi$. The HG
has a special place among the modified models. The field
equations in GR are differential equations of the second order,
thus, evading Ostrogradski instabilities arising
\cite{Ostrogradski,Woodard}. The HG is the most general variant of
the scalar-tensor theory of gravitation with motion equations of
the second order. The action density of HG contains several
functions that provide a broad phenomenology. This makes it
possible to solve important cosmological and astrophysical
problems (screening of the cosmological constant, kinetic
inflation, late de Sitter stage, hairy black holes, etc.)
\cite{Sushkov0, Appleby0, Sotiriou, Babichev, Maselli,
Muharlyamov, Bernardo0,Chien,Taniguchi}.

In this article we review the HG within the framework of the
BI cosmological model.  In the case  $G_{5X}\neq 0$ the
gravitational equations give consequences containing the nonlinear
anisotropic terms. Here we study the effects of nonlinear
anisotropy  with the function  $G_2(X,\phi)=-l\cdot \phi+A(X)$. Authors  \cite{Appleby, Bernardo} studied the linear potential
$l\cdot\phi$\,  for the isotropic Universe.  In the work
\cite{SushkovStar} the nonlinear anisotropy was studied for the
model
\begin{equation}
G_2=X-\Lambda\,, \, G_3=0\,, \, G_4=const\,,\,
G_5=const+\xi\sqrt{2X}\,. \label{}
\end{equation}
This model contains an early inflation and a late acceleration with suppressed
anisotropy. The anisotropy shows a maximum at intermediate times. In
the work \cite{Tahara} the nonlinearity leads to the effect of\,
"anisotropization"\, in the later times, that is, the Universe
evolves from an isotropic state to an anisotropic one.  The HG is not the only theory with the nonlinear anisotropy. In the paper \cite{Bhattacharya} the nonlinear growth of anisotropy in
BI spacetime in metric $f(R)$ cosmology was studied. Authors showed that any kind of anisotropy, if present initially, will be damped during the inflationary phase in quadratic gravity.

 We aim to obtain  and  investigate the  exact anisotropic cosmological solutions with certain properties. We want to find out what a model with the nonlinear anisotropic terms and the linear potential can offer. How interesting is this model for cosmology? What process will take place? Will there be a process of isotropization or \,"anisotropization"\,?

\section{Bianchi I model}\label{sec2}

 We  consider the homogeneous and anisotropic Bianchi I metric:
\begin{equation}
 ds^2 = -dt^2 + a^2_1(t)dx^2 + a^2_2(t)dy^2 + a^2_3(t)dz^2, \label{met0}
\end{equation}
with the three scale factors  $a_i$ and the scalar field $\phi(t)$ depending only
on $t$. Then the gravitational equations  take the form
\cite{SushkovStar}:
 $$G^0_0\left({\cal G}-2G_{4X}\dot{\phi}^2 -
2G_{4XX}\dot{\phi}^4 +2G_{5\phi}\dot{\phi}^2
+G_{5X\phi}\dot{\phi}^4\right) = G_2 - G_{2X}\dot{\phi}^2-
$$\begin{eqnarray} \label{00}- 3G_{3X}H\dot{\phi}^3 +
G_{3\phi}\dot{\phi}^2
  + 6G_{4\phi}H\dot{\phi} + 6G_{4X\phi}\dot{\phi}^3 H
 - 5G_{5X}H_1H_2H_3\dot{\phi}^3 - G_{5XX}H_1H_2H_3\dot{\phi}^5\,,\end{eqnarray}
$$ {\cal G}
G^{i}_{i}-(H_{j}+H_{k})\frac{d{\cal G}}{dt} = G_2 - \dot\phi
\frac{d G_3}{dt} +2 \frac{d}{dt}(G_{4\phi}\dot{\phi})
-$$\begin{eqnarray} \label{ii}-\frac{d}{dt}(G_{5X}\dot{\phi}^3
H_{j}H_{k})- G_{5X}\dot{\phi}^3 H_{j}H_{k}(H_{j}+H_{k}) \,.&&
\end{eqnarray}
Here the dot denotes  the $t$-derivative ($\dot{\phi}\equiv\frac{d\phi}{dt}$), one has $H_i=\dot {
a}_i/{ a}_i$, and the average Hubble parameter is
$H=\dfrac{1}{3}\sum\limits_{i=1}^3 H_i\equiv \dot{ a}/{ a}$ with
${ a}=({ a}_1{ a}_2{ a}_3)^{1/3}$. The Einstein tensor components
are
\begin{eqnarray}
&&G^0_0=-\left(H_1H_2 +H_2H_3 +H_3H_1\right)\,, \\
&&G^{i}_{i}=-\left(\dot{H}_{j} +\dot{H}_{k} +H_{j}^2 +H_{k}^2
+H_{j}H_{k}\right)\,,
\end{eqnarray}
where the triples of indices $\{i,j,k\}$ take values $\{1,2,3\}$,
$\{2,3,1\}$, or $\{3,1,2\}$. In addition, we define
\begin{equation}\label{def_G}
{\cal G} = 2G_4-2G_{4X}\dot{\phi}^2+G_{5\phi}\dot{\phi}^2\,.
\end{equation}

Varying the action (\ref{action}) with respect to $\phi$ yields the
equation
\begin{eqnarray}\label{scalar}
\frac{1}{\rm a^3}\frac{d}{dt}({\rm a^3} {\cal J})={\cal P}\,,
\end{eqnarray}
with $$ {\cal J} =\dot{\phi}\, \Big[ G_{2X}-2G_{3\phi}
+3H\dot\phi(G_{3X} -2G_{4X\phi})+$$
\begin{eqnarray} \label{J}+G^0_0(-2G_{4X}-2\dot\phi^2G_{4XX} +2G_{5\phi}
+G_{5X\phi}\dot\phi^2 )
 +H_1H_2H_3(3G_{5X}\dot{\phi} +G_{5XX}\dot{\phi}^3)\Big]\,,\end{eqnarray}
 $${\cal P} = G_{2\phi} -\dot{\phi}^2(G_{3\phi\phi}
+G_{3X\phi}\ddot \phi) +RG_{4\phi}+2G_{4X\phi}\dot\phi(3\ddot\phi
H-\dot\phi G^0_0)+ $$ \begin{eqnarray}
\label{P}+G^0_0G_{5\phi\phi}\dot{\phi}^2+G_{5X\phi}\dot{\phi}^3H_1H_2H_3\,.
\end{eqnarray}

 For convenience let's consider the following parametrization of three scalar factors:
\begin{equation}
a_1=ae^{\beta_{+}+\sqrt{3}\beta_{-}}\,,\, a_2=ae^{\beta_{+}-\sqrt{3}\beta_{-}} \,,\, a_3=ae^{-2\beta_{+}}\,,
\label{sdx}
\end{equation}
hence
\begin{equation}
ds^2 =
-dt^2+a^2(t)[e^{2(\beta_{+}+\sqrt{3}\beta_{-})}dx^2+e^{2(\beta_{+}-\sqrt{3}\beta_{-})}dy^2+e^{-4\beta_+}dz^2].
\label{met}
\end{equation}
This parametrization explicitly separates the isotropic and anisotropic parts.
The functions $e^{\beta_{+}+\sqrt{3}\beta_{-}}$\,,
$e^{\beta_{+}-\sqrt{3}\beta_{-}}$ and $e^{-2\beta_{+}}$ are the
deviation from isotropy, and $a(t)$ is the isotropic part. The
rate of expansion in the direction of $x$, $y$ and $z$ are given
by
\begin{equation}
H_1=H+\dot{\beta}_{+}+\sqrt{3}\dot{\beta}_{-}\,,\,
H_2=H+\dot{\beta}_{+}-\sqrt{3}\dot{\beta}_{-}\,, \,
H_3=H-2\dot{\beta}_{+}\,. \label{H}
\end{equation}
 The anisotropies are determined by $\dot\beta_{\pm}$, and if they vanish, then $H_1=H_2=H_3=H$ and the
Universe is isotropic.
This parametrization simplifies the process of integrating field equations.

We define the four arbitrary functions $G_i$ ($i=2,3,4,5$) as
follows
\begin{equation}\label{Gi} G_{2}=-l\cdot \phi+A(X)\,,\, G_3=G_3(X)\,,\, G_4=\frac{1}{16\pi}\,,\,  G_5=G_{5}(X)\,.\end{equation}
In the future,  the expression $l\cdot \phi$ will provide the
dynamic solution to $\dot\phi(t)$, $\dot{\beta}_{\pm}(t)$. The
theory with $G_5(X)$ gives a nontrivial behavior of anisotropy.
Taking into account (\ref{met}), (\ref{H}) and (\ref{Gi}) from
equations (\ref{00}), (\ref{ii}) and (\ref{scalar}) we obtain the
consequences
 $$\frac{3}{8\pi}\big(H^2-\sigma^2\big) =l\phi-A +\dot{\phi}^2A_{X}+ 3G_{3X}H\dot{\phi}^3
 +$$
\begin{eqnarray}
\label{tad00}+\dot{\phi}^3(5G_{5X} +G_{5XX}\dot{\phi}^2) (H -
2\dot{\beta}_{+}) \big[(H+\dot{\beta}_+)^2
-3\dot{\beta}_{-}^2\big]\,,
\end{eqnarray}
$$\frac{1}{8\pi}\big(2\dot{H}+3H^2 + 3\sigma^2\big) =l\phi -A
+G_{3X}\dot{\phi}^2\ddot{\phi}+
 $$
\begin{eqnarray}+\frac{d}{dt}\left[G_{5X}\dot{\phi}^3
\left(H^2-\sigma^2\right)\right]
+2G_{5X}\dot{\phi}^3\left(H^3+\dot{\beta}_{+}^3-3\dot{\beta}_{+}\dot{\beta}_{-}^2
\right),  \label{dh} &&
\\
\label{bb} \frac{\dot{\beta}_{+}}{8\pi} +G_{5X}\dot\phi^3
\left(\dot\beta_{-}^2-\dot\beta_{+}^2 -H\dot\beta_{+}\right) =
\frac{C_+}{{ a}^3}, &&
\\
\label{bbb} \frac{\dot{\beta}_{-}}{8\pi} +G_{5X}\dot\phi^3
\left(2\dot\beta_{+}\dot\beta_{-} -H\dot\beta_{-}\right) =
\frac{C_-}{{ a}^3}, &&
\end{eqnarray}
$$\dot{\phi}\, \Big[ A_{X}+3HG_{3X}
\dot\phi
+(H-2\dot\beta_+)[(H+\dot\beta_+)^2-3\dot\beta_-^2]\times$$
\begin{eqnarray}  \label{scalartad00}
 \times(3G_{5X}\dot{\phi}
+G_{5XX}\dot{\phi}^3)\Big]=\frac{C_\phi}{{ a}^3}-\frac{l}{a^3}\int
a^3(t)dt, \end{eqnarray} where $C_\phi$, $C_+$ and $C_-$ are
integration constants. The constants $C_{\pm}$ correspond to the
anisotropic charges.  For convenience, we introduce
\begin{equation}\label{sigma}
\sigma^2 = \dot{\beta}^2_{+} + \dot{\beta}^2_{-}\,.
\end{equation}
The theory with $G_{5X}\neq0$ gives the system of nonlinear equations
(\ref{bb}), (\ref{bbb}) for $\dot\beta_{\pm}$. From this point of
view, we consider the nonlinear anisotropic model.

Further, we put
\begin{equation}C_\phi=C_-=C_+=0\,.\end{equation}
Then the simplest solution is the isotropic one,
\begin{equation}\label{sigma}
\dot{\beta}_{\pm}=0\,.
\end{equation}
In addition, since the equations are nonlinear, there are also solutions with $\dot{\beta}_{\pm}\neq
0$:
\begin{equation}\label{b0}\dot\beta_{+}=\frac{1}{2}\left(H-\frac{1}{8\pi \cdot
G_{5X}\dot{\phi}^3}\right),\,\,
\dot\beta_{-}=\pm\frac{\sqrt{3}}{2}\left(H-\frac{1}{8\pi \cdot
G_{5X}\dot{\phi}^3}\right)\,,\end{equation}
\begin{equation}\label{sigmab0}\sigma^2=\left(H-\frac{1}{8\pi \cdot G_{5X}\dot{\phi}^3}\right)^2.\end{equation}
In the solution (\ref{b0}), the signs "$+$" and "$-$" of
$\dot\beta_{-}$ correspond to
\begin{equation}\label{pl}(+): \,H_1=3H-2\cdot\frac{1}{8\pi \cdot G_{5X}\dot{\phi}^3}\,,\, H_2=H_3=\frac{1}{8\pi \cdot G_{5X}\dot{\phi}^3}\,,\end{equation}
\begin{equation}\label{min}(-): \,H_2=3H-2\cdot\frac{1}{8\pi \cdot G_{5X}\dot{\phi}^3}\,,\, H_1=H_3=\frac{1}{8\pi \cdot G_{5X}\dot{\phi}^3}\,.\end{equation}
Thus, the assumption $C_-=C_+=0$ gives the locally rotationally
symmetric (LRS) BI model. We will consider the model
(\ref{pl}).

In view of (\ref{b0}) and (\ref{sigmab0}), from equations
(\ref{tad00}) and (\ref{scalartad00}) we obtain
$$3H\left\{G_{3X}\dot{\phi}^3+\frac{1}{(8\pi)^2 \cdot G_{5X}\dot{\phi}^3}\left[3+\frac{G_{5XX}\cdot\dot{\phi}^2}{G_{5X}}\right]\right\}
=$$\begin{equation}\label{G5grav}=-l\phi+A-\dot{\phi}^2A_{X}+\frac{1}{(8\pi)^3
\cdot(
G_{5X}\dot{\phi}^3)^2}\left[7+\frac{2G_{5XX}\cdot\dot{\phi}^2}{G_{5X}}\right],\end{equation}
$$3H\left\{G_{3X}\dot{\phi}^3+\frac{1}{(8\pi)^2 \cdot G_{5X}\dot{\phi}^3}\left[3+\frac{G_{5XX}\cdot\dot{\phi}^2}{G_{5X}}\right]\right\}
=$$\begin{equation}\label{G5scalar}=-\dot{\phi}^2A_{X}-\frac{l\cdot\dot{\phi}}{a^3}\int
a^3dt+\frac{2}{(8\pi)^3 \cdot(
G_{5X}\dot{\phi}^3)^2}\left[3+\frac{G_{5XX}\cdot\dot{\phi}^2}{G_{5X}}\right].\end{equation}
The  equation (\ref{dh}) can be ignored, since it is automatically
fulfilled by virtue of the Bianchi identities. The combination of
(\ref{G5grav}) and (\ref{G5scalar}) gives the equation:
\begin{equation}\label{diffrenc}\frac{1}{(8\pi)^3 \cdot (G_{5X}\dot{\phi}^3)^2}-l\phi+A=-\frac{l\cdot\dot{\phi}}{a^3}\int
a^3dt\,.\end{equation} Let us note an important property of the
presented model. If $l=0$ then the system (\ref{b0}),
(\ref{G5grav}), (\ref{G5scalar}) can only have the stationary
solution $\dot\beta_{\pm}$, $H$, $\dot\phi = const$. A necessary
condition for the existence of the dynamic solution
$\dot\beta_{\pm}(t)$, $\dot\phi(t)$ is the presence of the term
$l\cdot\phi\neq 0$. In the work \cite{SushkovStar} the dynamics
of the solution was provided by nonzero charges $C_\phi$,
$C_{\pm}$.

Next, we use the reconstruction method. Let's make an assumption for
the average scale factor
\begin{equation}\label{}\frac{l}{a^3}\int
a^3dt=\mu=const>0 \,.\end{equation}
This assumption gives the exact solution, and in this case the Universe is expanding with acceleration:
\begin{equation}\label{Sitt} H=\frac{l}{3\mu}=const\,, \, a(t)=a_*\exp{\left(\frac{l\cdot t}{3\mu}\right)}\,, \, l>0\,.\end{equation}
In other words, we consider the isotropization process  of
the Universe, which is approaching de Sitter's world. The proposed model can describe the primary inflation of the early Universe. The type of the scale factor (\ref{Sitt}) is substantiated and used by many authors to describe the cosmic inflation
in anisotropic models \cite{Bhattacharya,Mandal,Rodrigues, Akarsuaaa}. The model (\ref{Sitt}) is interesting in the context of inflationary mechanism for suppressing the anisotropy.
We choose
the function $A(X)$ as follows
\begin{equation}\label{Ax} A(X)=-\mu\dot{\phi}=-\mu\sqrt{2X}\,,\, \dot{\phi}\geq0\,.\end{equation}
The model with $G_2\propto \sqrt{X}$ corresponds to {\it Cuscuton}
scenarios \cite{Afshord, Quintin0, Quintin1}. From the equation
(\ref{diffrenc}) it follows
\begin{equation}\label{phe}\frac{1}{(8\pi)^3 \cdot (G_{5X}\dot{\phi}^3)^2}-l\cdot\phi=0
\Rightarrow \frac{1}{8\pi \cdot G_{5X}\dot{\phi}^3}=\pm\sqrt{8\pi
l\cdot\phi}\,.\end{equation} We choose the "+" sign. In this case
the Universe expands in all directions, $H_i>0$ (see (\ref{pl}),
(\ref{min})).

We want to get a model that is isotropic in later times
($t\rightarrow +\infty$). Therefore, it must be fulfilled
\begin{equation}\frac{\dot\beta_{\pm}}{H}\rightarrow 0\,\,\, \text{as}\,\, \, t\rightarrow +\infty\,,\end{equation}
i.e
\begin{equation}\label{izotrop}\frac{1}{8\pi \cdot G_{5X}\dot{\phi}^3}\rightarrow H\,\,\, \text{as}\,\, \, t\rightarrow +\infty\,.\end{equation}
\begin{figure}[h]
\includegraphics[width=8cm]{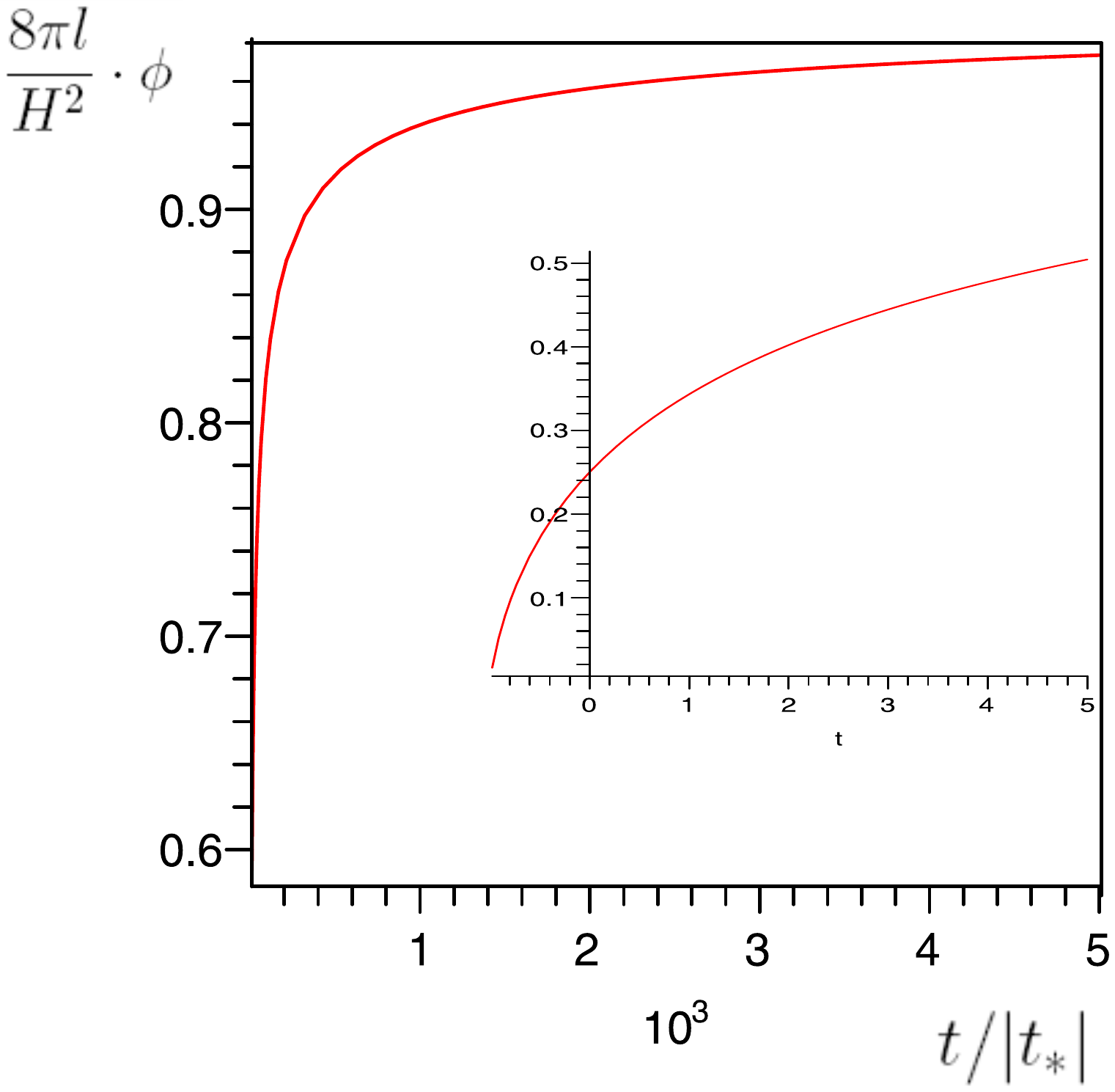}
\caption{The profile of the scalar field $\phi$. \label{scalar1_2}}
\end{figure}
The dynamics of $\dot\phi(t)$ allows one to construct solution
(\ref{b0}) with the isotropization. To satisfy the isotropization
condition (\ref{izotrop}), we put
\begin{equation} \label{dds} \frac{1}{8\pi \cdot G_{5X}\dot{\phi}^3}=H+\gamma \dot{\phi}^{1/3}\,, \, \gamma=const\,.\end{equation}
Then the equation (\ref{phe}) is transformed to the form
\begin{equation}H+\gamma
\dot{\phi}^{1/3}=\sqrt{8\pi l\cdot\phi}\,.\end{equation} This
equation has a nonsingular solution
\begin{equation}\label{pheSol} \phi=\frac{H^2}{8\pi l}\cdot\frac{1+\frac{8\pi l H }{|\gamma|^3}t}{\left(1+ \sqrt{1+\frac{8\pi l H }{|\gamma|^3}t}\,\right)^2}
\,,\, \gamma<0\,,\,  t\geq t_*\equiv-\frac{|\gamma|^3}{8\pi l
H}\,.\end{equation} The scalar field $\phi$ is the
bounded function, $0\leq\phi<H^2/(8\pi l)$.  As seen in Fig.\ref{scalar1_2},  the function $\phi$
tends monotonically to a constant value over time and there is no infinite discontinuity. With time the linear potential $V=l\cdot \phi$ begins to behave like the cosmological constant.
\begin{figure}[h]
\includegraphics[width=8cm]{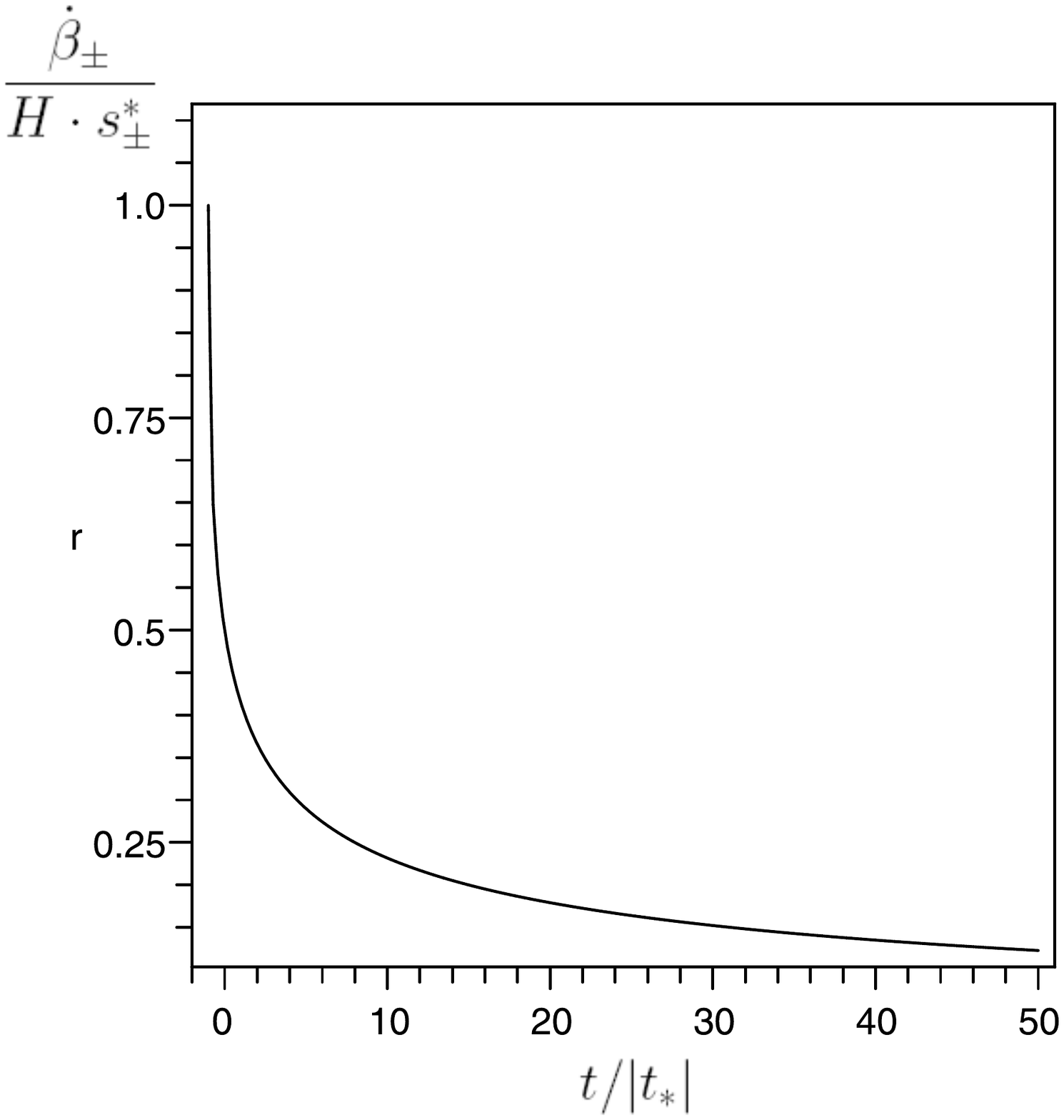}
\caption{Anisotropy profiles $\dot\beta_{\pm}$. \label{aniz}}
\end{figure}
Considering (\ref{b0}), (\ref{dds}) and (\ref{pheSol}) we get
$$\frac{\dot\beta_{\pm}}{H}=\frac{s^*_\pm}{1+
\sqrt{1+\frac{t}{|t_*|}}}\rightarrow 0 \,\,\, \text{as}\,\, \,
t\rightarrow +\infty\,; s^*_+=1/2\,, s^*_-=\sqrt{3}/2\,;
$$\begin{equation}\label{bbett}\frac{\dot\beta_{\pm}}{H}\rightarrow s^*_\pm \,\,\, \text{as}\,\,
\, t\rightarrow t_*\,.\end{equation}  As seen in Fig.\ref{aniz}, the functions $\dot\beta_{\pm}$
tend monotonically to zero over time, that is,  the Universe becomes
isotropic at later times. The functions $\dot\beta_{\pm}$ are
finite at the beginning ($t_*$) of the Universe. This is a
non-standard behavior of the  Universe anisotropy.

\begin{figure}[h]
\includegraphics[width=8cm]{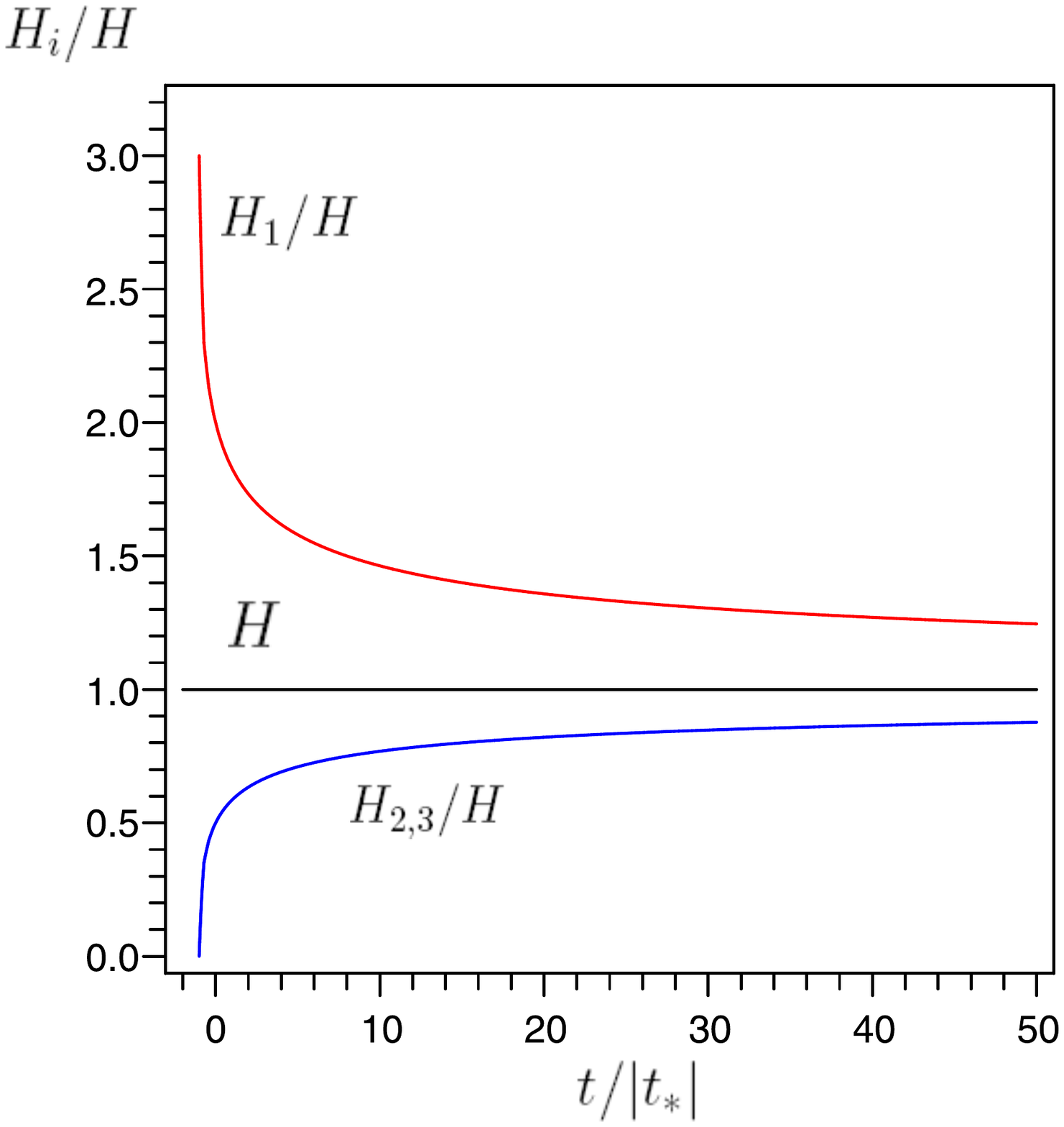}
\caption{Hubble parameter profiles $H_i$. \label{habl}}
\end{figure}

The Universe is expanding along the $x$, $y$ and $z$ axes:
\begin{equation}\label{h1h2h3}H_1=H\cdot\frac{3+ \sqrt{1+\frac{t}{|t_*|}}}{1+
\sqrt{1+\frac{t}{|t_*|}}}>0 \,,\, H_2=H_3=H\cdot \frac{
\sqrt{1+\frac{t}{|t_*|}}}{1+ \sqrt{1+\frac{t}{|t_*|}}}>0
\,.\end{equation}   As seen in
Fig.\ref{habl}, the functions $H_i$ tend monotonically to a constant value over time, $H_i\rightarrow H$=const, $t/t_*\gg 1$.  There is no infinite discontinuity. The Universe is approaching the de Sitter’s world.

 We can find $\beta_{\pm}$ from the formula (\ref{bbett}):
\begin{equation}\label{betq} \beta_{\pm}=2s^*_\pm H |t_*|\left[\sqrt{1+\frac{t}{|t_*|}}-
\ln\left(1+\sqrt{1+\frac{t}{|t_*|}}\,\right)\right]+const_{\pm}   \,.\end{equation}
On integration (\ref{h1h2h3}) we get  the scale factors
$$a_1(t)=a_1^*\exp \left\{H|t_*|\left[1+\frac{t}{|t_*|}+4\sqrt{1+\frac{t}{|t_*|}}-4\ln\left(1+\sqrt{1+\frac{t}{|t_*|}}\,\right)\right]\right\}\,,$$
\begin{equation}a_{2,3}(t)=a_{2,3}^*\exp
\left\{H|t_*|\left[1+\frac{t}{|t_*|}-2\sqrt{1+\frac{t}{|t_*|}}+2\ln\left(1+\sqrt{1+\frac{t}{|t_*|}}\,\right)\right]\right\}\,,\,
t\geq t_*\,,\end{equation} where $a_1^*$ and $a_{2,3}^*$ are
integration constants, $a_i^*=a_i(t_*)$.   This cosmological model
does not contain a singular point. The Ricci scalar (invariant)
$$R=2\left[\frac{\ddot{a}_1}{a_1}+\frac{\ddot{a}_2}{a_2}+\frac{\ddot{a}_3}{a_3}+
\frac{\dot{a}_1\dot{a}_2}{a_1a_2}+\frac{\dot{a}_1\dot{a}_3}{a_1a_3}+\frac{\dot{a}_2\dot{a}_3}{a_2a_3}\right]=$$
\begin{equation}=\frac{6H^2}{\left(1+
\sqrt{1+\frac{t}{|t_*|}}\,\right)^2} \cdot
\left(5+4\cdot\sqrt{1+\frac{t}{|t_*|}}+\frac{2t}{t_*}\right)\end{equation}
has no singularities.

Using (\ref{Sitt}) and (\ref{Ax}) from the equation
(\ref{G5grav}), we restore the function $G_3(X)$:
\begin{equation}G_{3X}=\frac{1}{(8\pi)^2\cdot G_{5X}\dot{\phi}^6}\left(3+\frac{G_{5XX}\cdot\dot{\phi}^2}{G_{5X}}\right)\left(-1+
\frac{1}{12H\pi\cdot G_{5X}\dot{\phi}^3}\right)\,,\end{equation}
then
\begin{equation} G_3=\frac{\gamma}{48\sqrt[3]{2}\,\pi}\left[-X^{-1/3}+\frac{4\sqrt{2}\,\gamma}{HX^{1/6}}\right]+const\,.\end{equation}
From the assumption (\ref{dds}), we restore the function $G_5(X)$:
$$G_5=-\frac{3|\gamma|^2}{8\cdot 2^{1/6}\pi H^3 X^{1/6}}-\frac{3|\gamma|}{16\cdot 2^{1/3}\pi H^2 X^{1/3}}-
\frac{1}{8\cdot 2^{1/2}\pi H
X^{1/2}}+$$\begin{equation}+\frac{3|\gamma|^3}{8\pi
 H^4}\ln\left|\frac{2^{1/6}|\gamma|X^{1/6}}{2^{1/6}|\gamma|X^{1/6}-H}\right|+const\,.\end{equation}
Thus, we have presented the theory that allows the isotropization process of the Universe with the finite anisotropy.

\section{Conclusions}

 We have studied the anisotropic BI cosmology within the HG with functions: $G_2(X,\phi)=-l\cdot \phi+A(X)$, $G_3(X)\neq 0$, $G_4=const$ and $G_5(X)\neq0$. Our aim was to see what the nonlinear anisotropy with the
linear potential could produce.  We have studied the isolated influence of the linear potential on the dynamics of the field functions, assuming that the scalar charge and the anisotropic charges are equal to zero, $C_\phi=C_{\pm}=0$. The value of anisotropic charges affects the symmetry of space-time.  The assumption $C_{\pm}=0$ gives the locally rotationally symmetric Bianchi I model: $a_1\neq a_2=a_3$.

 We have presented the exact cosmological solution.
It has the following properties. First, the anisotropic terms are always finite and they reach their a maximum value at the initial moment. This is the non-standard behavior of the  Universe anisotropy. Secondly, the anisotropy suppression occurs
during the inflationary stage,  and it approaches
zero at later times. When the model is isotropized over time, we get the de Sitter’s world. Third, the cosmological model does
not contain a singular point.

 To find the exact solution, we applied the reconstruction method. This method is often used in the modified
theories of gravity \cite{Esposito, Muharlyamov, Bernardo0, Appleby, Bernardo}.  The reconstruction method is interesting and optimal for assessing the viability of the modified theory of gravity under study. The Lagrangian functions $G_2$, $G_4$ are entered manually. A priori, the average Hubble parameter was set constant, $H=const$. Modeling the process of isotropization, we have chosen the function $G_{5X}$. Such assumptions allow you to restore the desired functions $G_3$, $G_5$.

\acknowledgments This work is supported by the Russian Foundation
for Basic Research (Grant No. 19-52-15008).

\end{document}